\documentclass[12pt,epsf]{article}
\usepackage{graphicx}
\begin{document}
%%%%%%%%%%%%%%%%%%%%%%%%%%%%%%%%%%%%%%%%%%%

\def\a{\alpha}
\def\b{\beta}
\def\c{\varepsilon}
\def\d{\delta}
\def\e{\epsilon}
\def\f{\phi}
\def\g{\gamma}
\def\h{\theta}
\def\k{\kappa}
\def\l{\lambda}
\def\m{\mu}
\def\n{\nu}
\def\p{\psi}
\def\q{\partial}
\def\r{\rho}
\def\s{\sigma}
\def\t{\tau}
\def\u{\upsilon}
\def\v{\varphi}
\def\w{\omega}
\def\x{\xi}
\def\y{\eta}
\def\z{\zeta}
\def\D{\Delta}
\def\G{\Gamma}
\def\H{\Theta}
\def\L{\Lambda}
\def\F{\Phi}
\def\P{\Psi}
\def\S{\Sigma}

\def\o{\over}
\def\beq{\begin{eqnarray}}
\def\eeq{\end{eqnarray}}
\newcommand{\gsim}{ \mathop{}_{\textstyle \sim}^{\textstyle >} }
\newcommand{\lsim}{ \mathop{}_{\textstyle \sim}^{\textstyle <} }
\newcommand{\vev}[1]{ \left\langle {#1} \right\rangle }
\newcommand{\bra}[1]{ \langle {#1} | }
\newcommand{\ket}[1]{ | {#1} \rangle }
\newcommand{\EV}{ {\rm eV} }
\newcommand{\KEV}{ {\rm keV} }
\newcommand{\MEV}{ {\rm MeV} }
\newcommand{\GEV}{ {\rm GeV} }
\newcommand{\TEV}{ {\rm TeV} }
\def\diag{\mathop{\rm diag}\nolimits}
\def\Spin{\mathop{\rm Spin}}
\def\SO{\mathop{\rm SO}}
\def\O{\mathop{\rm O}}
\def\SU{\mathop{\rm SU}}
\def\U{\mathop{\rm U}}
\def\Sp{\mathop{\rm Sp}}
\def\SL{\mathop{\rm SL}}
\def\tr{\mathop{\rm tr}}

\def\IJMP{Int.~J.~Mod.~Phys. }
\def\MPL{Mod.~Phys.~Lett. }
\def\NP{Nucl.~Phys. }
\def\PL{Phys.~Lett. }
\def\PR{Phys.~Rev. }
\def\PRL{Phys.~Rev.~Lett. }
\def\PTP{Prog.~Theor.~Phys. }
\def\ZP{Z.~Phys. }

%%%%%%%%%%%%%%%%%%%%%%%%%%%%%%%%%%%%%%%%%%%%%%%%%%%%%%%%%%%%%%%%%%%%

\baselineskip 0.7cm

\begin{titlepage}

\begin{flushright}
SLAC-PUB-12875\\
UT-07-31
\end{flushright}

\vskip 1.35cm
\begin{center}
\center{
{\large \bf
$R$-invariant New Inflation Model\\ vs\\ Supersymmetric Standard Model
}}
\vskip 1.2cm
M. Ibe${}^{1}$ and  Y.  Shinbara${}^{2}$
\vskip 0.4cm
${}^{1}$
{\it Stanford Linear Accelerator Center, Stanford University,
                Stanford, CA 94309 and } \\
{\it Physics Department, Stanford University, Stanford, CA 94305}\\
${}^2${\it Department of Physics, University of Tokyo, Tokyo 113-0033, Japan}

\vskip 1.5cm
\abstract{
We revisit the implications of the $R$-invariant New Inflation model to 
the supersymmetric standard model in light of recent discussion of gravitino production processes
by the decay of the inflaton or the supersymmetry breaking field.
We show that the models with supergravity mediation do not go well with
the $R$-invariant New Inflation model, where the gravitino abundance produced 
by the decay of the inflaton or the supersymmetry breaking field 
significantly exceeds the bounds from cosmological observations without fine-tuning.
We also show that  the models with gauge mediation can go together with $R$-invariant New 
Inflation model, 
where the dark matter and the baryon asymmetry are consistently explained without 
severe fine-tuning.
 }
\end{center}
\end{titlepage}

\setcounter{page}{2}

\section{Introduction}
The Supersymmetric Standard Model (SSM) is considered as one of the most promising candidates for
physics beyond the Standard Model (SM), which will be tested at the coming Large Hadron Collider (LHC) experiments.
Once the supersymmetric particles are discovered, the next important task will be to determine how the supersymmetry (SUSY) breaking occurs and how the breaking effects are mediated to the SSM sector.
So far, variety of mediation mechanisms have been proposed, and they are roughly classified 
into three categories.
The first class is called models with ``gravity mediation (SUGRA)'', 
where the communications between a SUSY breaking sector and 
the SSM sector are suppressed by the Planck scale 
($M_{\rm PL}$)~\cite{Chamseddine:1982jx,Hall:1983iz}.
The second class is called models with ``gauge mediation (GMSB)'',
 where the breaking effects are mediated at the 
lower energy scale than $M_{\rm PL}$ via gauge interactions of the 
SSM~\cite{Dine:1981za,Dine:1993yw,Dine:1994vc,Dine:1995ag}. 
The final class is models with ``anomaly mediation (AMSB)'' in which the breaking effects mediated to the SSM sector are suppressed by more than $M_{\rm PL}$~\cite{Randall:1998uk,Giudice:1998xp}.
Since the characteristic scale of the SUSY breaking (or the size of the gravitino mass) is different among the above categories, the SUSY breaking scale (or of the gravitino mass) can represent 
the mediation mechanisms.

Fortunately, there are already some evidences that constrain the size of the gravitino mass 
from cosmology.
For example, the late time decay of the unstable gravitino produced after inflation may spoil 
the success of the  Big Bang Nucleosynthesis (BBN) depending on the reheating temperature of
the universe $T_{R}$ 
(for recent works, see \cite{Kawasaki:2004qu,Jedamzik:2006xz} and reference therein). 
On the other hand, the abundance of stable gravitino is also constrained not to exceed the observation 
of the dark matter density~\cite{Moroi:1993mb,Bolz:2000fu,Pradler:2006qh,Rychkov:2007uq}.
Furthermore, recent works on the gravitino abundance produced by
the decay of moduli~\cite{Endo:2006zj,Nakamura:2006uc,Dine:2006ii,Endo:2006tf} 
and inflatons~\cite{Kawasaki:2006gs,Asaka:2006bv,Endo:2006qk,Endo:2006xg,Endo:2007ih} 
have shown that there are much more sever constraints on the gravitino mass 
depending on models of inflation. 

In this paper, we further pursue the constrains on the mediation mechanisms 
(i.e. the sizes of the gravitino mass) based on a class of New Inflation model which is dubbed $R$-invariant New Inflation model~\cite{Kumekawa:1994gx,Izawa:1996dv}.
The $R$-invariant New Inflation model has many attractive features.
First attractive feature is the simpleness of the model.
The model consists of only one chiral-sueprfield,
and the inflation dynamics are determined by only three parameters.
Another attractive feature is that it predicts the spectral index $n_{s}$ of the cosmic microwave background radiation as $n_{s}\simeq 0.95$ in a large parameter space~\cite{Izawa:2003mc,Ibe:2006fs}, 
which is well consistent with the WMAP observation~\cite{Spergel:2006hy}.
Finally, the most interesting feature from the viewpoint of the SSM model building 
is that the gravitino mass is determined by the energy scale of the inflation, 
i.e. the Hubble parameter during inflation.

In Ref.~\cite{Ibe:2006fs}, we showed that the $R$-invariant New Inflation model is well compatible with the SUGRA models (i.e. $m_{3/2}=O(1)$\,TeV), 
while providing the right amount of the baryon asymmetry of the universe 
by leptogenesis~\cite{Fukugita:1986hr} via the decay of the inflaton into right-handed 
(s)neutrinos~\cite{Campbell:1992hd,Kumekawa:1994gx,Asaka:1999yd,Ibe:2005jf}.
In Ref.~\cite{Ibe:2006am}, we also showed that the $R$-invariant New Inflation model 
eludes the Polonyi-induced gravitino problem in the SUGRA models.
As we will show, however,  such compatibility
with the SUGRA model is tainted by a large amount of gravitino produced by 
the decay of the inflaton or the SUSY breaking field unless we require, 
which cannot be avoided without fine-tuning.
On the other hand, we also show that the $R$-invariant New Inflation model can go well with
GMSB models even if we take into account of the gravitino production by the decay of 
the inflaton or SUSY breaking fields.

The construction of this paper is as follows. 
We  summarize relevant features of the $R$-invariant New Inflation model in the next section.
In section 3, we study the consistency of the inflation model with the SSM based on the SUGRA model,
in light of the gravitino production from the decay of the inflaton or the SUSY breaking field.
In section 4, we study the gravitino production for the gravitino mass scale characteristic for
GMSB.

%%%%%%%%%%%%%%%%%%%%%%%%%%%%%%%%%%%
\section{$R$-invariant New Inflation model}\label{sec:newinflation}
Let us summarize the $R$-invariant New Inflation model 
considered in Ref.~\cite{Kumekawa:1994gx,Izawa:1996dv}. 
The model is defined by the following superpotential and K\"ahler potential of an inflaton chiral superfield $\phi$,
\begin{eqnarray}
W_{\rm inf} = v^{2} \phi - \frac{g}{n+1}\phi^{n+1},
\label{eq:Super}
 \end{eqnarray}
and 
\begin{eqnarray}
K_{\rm inf} = |\phi|^{2} + \frac{k}{4} |\phi|^{4}+ \cdots.
\label{eq:Kahler}
\end{eqnarray}
Here,  $v^{2}$ denotes a dimensionful parameter,  $g$ and $k$  dimensionless coupling constants,
and $n$ is integer.
We can take the parameters $v^{2}$ and $g$ positive without loss of  generality. 
Hereafter, we take the unit where the reduced Planck scale, $M_{\rm PL}\simeq 2.4\times 10^{18}$ GeV,
equals to one unless we specify. 
The above superpotential is generic under a discrete $Z_{2n}$\,$R$-symmetry with $\phi$'s charge 2.

By taking account of supergravity effects, the effective scalar potential of the inflaton 
$\varphi = \sqrt{2} {\rm Re}[\phi]$ is well approximated by
\begin{equation}
 V(\varphi) \simeq v^4 - \frac{k}{2}v^4 \varphi^2
 -\frac{g}{2^{\frac{n}{2}-1}}v^2\varphi^n
 +\frac{g^2}{2^n}\varphi^{2n},
\label{eq:potential}
\end{equation}
during inflationary period (i.e. $\varphi\sim 0$).
The potential becomes very flat for $n\geq3$ and $|k|\ll 1$, and it serves as a New Inflation potential
with the Hubble parameter $H_{\rm inf} \simeq v^{2}/3$ for $k>0$.
From the COBE normalization of the amplitude of the primordial density fluctuation, the Hubble parameter can be expressed as a function of $g$ for $k\lsim 10^{-2}$,
\begin{eqnarray}
\label{eq:cobe4}
H_{\rm inf} &\simeq& 10^{5.4}\,{\rm GeV} \times \frac{1}{g},\, (n=4),\\
\label{eq:cobe5}
H_{\rm inf} &\simeq& 10^{8.6}\,{\rm GeV} \times \frac{1}{g^{1/2}},\, (n=5),\\
\label{eq:cobe6}
H_{\rm inf} &\simeq& 10^{9.9}\,{\rm GeV} \times \frac{1}{g^{1/3}},\, (n=6),
\end{eqnarray}
and $H_{\rm inf}$ increases for larger $n$.
Here, we are also assuming that the e-folding number $N_{e}$ at the horizon crossing 
to be 50, although our discussion barely depends on this assumption as long as $N_{e}=O(10)$.
The dependence of $H_{\rm inf}$ on $k$ is also weak as long as $k\lsim 10^{-2}$ 
(see Refs.~\cite{Ibe:2006fs} for details ).

The remarkable feature of the present model is that the inflation scale $H_{\rm inf}$ (or $v$) is
directly related to the gravitino mass~\cite{Kumekawa:1994gx}.
As we see from Eq.~(\ref{eq:Super}), the superpotential develops a non-vanishing vacuum expectation value (VEV), i.e. $\vev {W_{\rm inf}}\neq 0$, once the inflaton settles to its VEV at $\phi_{0}\simeq (v^{2}/g)^{1/n}$ after inflation.
On the other hand, we cannot introduce a large constant term in the superpotential,
since a constant term of $O(\vev{W_{\rm inf}})$ results in a small e-folding number, 
$N_{e}\ll O(10)$.
Thus, we have no free parameter for the VEV of the total superpotential,
and the gravitino mass is given by $\vev {W_{\rm inf}}$,
\begin{equation}
 m_{3/2} = \vev{W_{\rm inf}} \simeq \frac{n v^2}{n+1}  \left(\frac{v^2}{g}\right)^{\frac{1}{n}}.
  \label{eq:gmass}
\end{equation}
Therefore, the gravitino mass has an one-to-one correspondence with the 
Hubble parameter, $H_{\rm inf}= v^{2}/3$ (see Eqs.~(\ref{eq:cobe4})--(\ref{eq:cobe6})).

\begin{figure}[t]
 \begin{minipage}{.48\linewidth}
  \includegraphics[width=.9\linewidth]{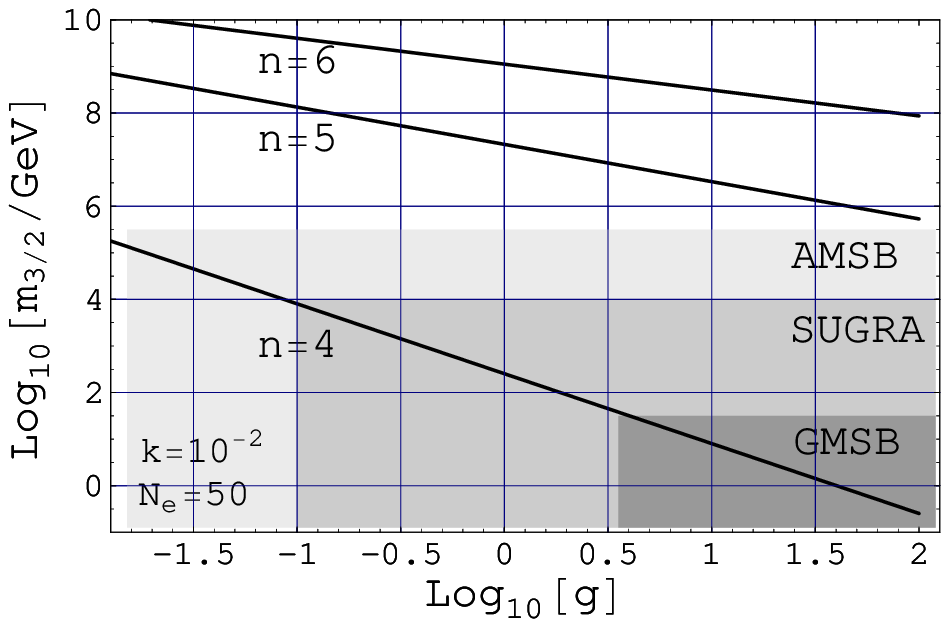}
 \end{minipage}
 \begin{minipage}{.50\linewidth}
  \includegraphics[width=.9\linewidth]{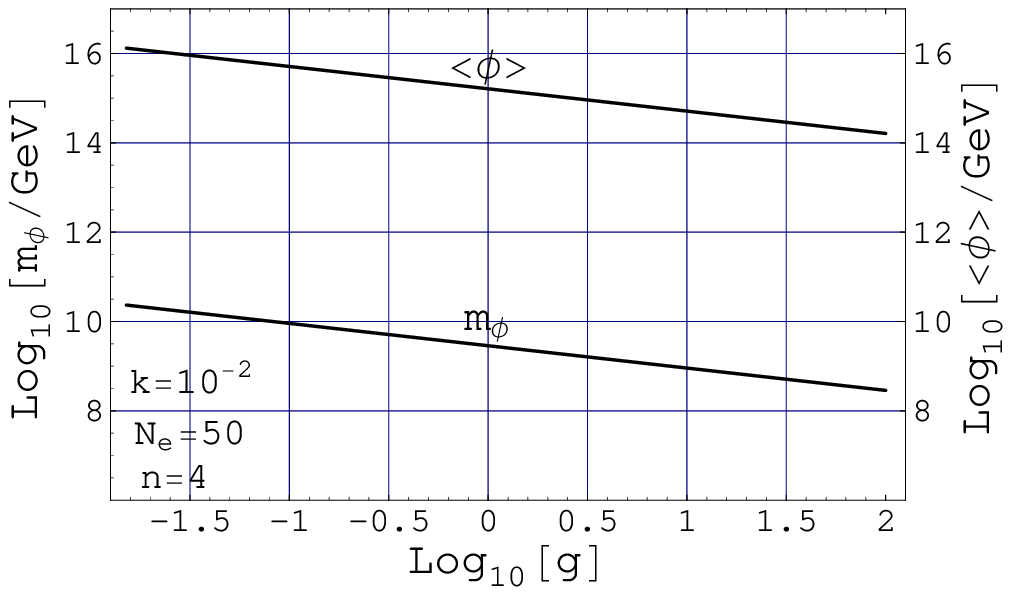}
 \end{minipage}
\caption{Left) The $g$ dependence of the gravitino mass for a given value of $n$.
The shaded region represents the typical gravitino mass regions for 
GMSB ($m_{3/2}\lsim 30$\,GeV), SUGRA ($m_{3/2}= O(100)\,{\rm GeV}-O(1)$\,TeV),
and AMSB ($m_{3/2}=O(10)-O(100)$\,TeV).
Right) The $g$ dependences of the mass and the VEV of the inflaton for $n=4$.
}
\label{fig:gravitino}
\end{figure}

The left panel of Fig.~\ref{fig:gravitino} shows the $g$ dependence of the gravitino mass 
for a given value of $n$.
From the figure, we see that the predicted gravitino mass for $n\geq 5$ is too large for 
all mediation mechanisms listed above, 
while the gravitino mass for $n=4$ is compatible with all the three mediation mechanisms.
Notice that it is rather difficult to obtain the spectral index $n_{s}$ which is consistent 
with the observed spectral index, $n_{s}= 0.951^{+0.015}_{-0.019}$~\cite{Spergel:2006hy} for $n=3$. 
Thus, we do not pursue the case with $n=3$ further.
For these reasons, we concentrate on the case of $n=4$ in the following argument, where the gravitino
mass can be well approximated by,
\begin{eqnarray}
m_{3/2} \simeq 300 \,{\rm GeV}
\times \frac{1}{g^{3/2}}.
\end{eqnarray}

%%%%%%%%%%%%%%%%%%%%%%%%%%%%%%%%%%%%%%%%%%%%%
%%%%%%%%%%%%%%%%%%%%%%%%%%%%%%%%%%%%%%%%%%%%%

In the right panel of Fig.~\ref{fig:gravitino}, 
we also plot the $g$ dependences of the mass and the VEV of the inflaton field $\phi$ for $n=4$
which are given by,
\begin{equation}
\label{eq:phimass}
m_{\phi} \simeq n g \phi_{0}^{n-1} \simeq n v^{2} \left(\frac{v^2}{g}\right)^{-\frac{1}{n}}
\simeq 3\times 10^{9}\,{\rm GeV}\times \frac{1}{\sqrt g},
\end{equation}
\begin{equation}
\vev\phi = \frac{1}{\sqrt{2}}\varphi_{0}\simeq \left(\frac{v^2}{g}\right)^{\frac{1}{n}}
\simeq 2\times 10^{15}\,{\rm GeV}\times \frac{1}{\sqrt g},
 \label{eq:phivev}
\end{equation}
respectively.
From the figure, we see that $m_{\phi}\simeq 10^{8-10}$\,GeV and $\vev \phi=10^{14-16}$\,GeV
for a wide range of parameter space.

Before closing this section, we comment on the possible range of the parameter $g$.
Since the K\"ahler potential receives radiative corrections, we need to require at least
$g<O(10)$ to keep perturbativity of the model.
Thus, in the following argument, we simply assume $g\lsim 10$ which corresponds to,
\begin{eqnarray}
 m_{3/2}\gsim 10\,{\rm GeV}.
\end{eqnarray}
We should, however, keep in mind that we need some degree of fine-tuning 
between the tree level contribution and the radiative corrections to the 
quartic coupling in the K\"ahler potential in Eq.~(\ref{eq:Kahler})
to keep the effective quartic coupling small, 
i.e. $|k|\lsim 10^{-2}$, when the coupling constant $g$ is $O(1)$.

%%%%%%%%%%%%%%%%%%%%%%%%%%%%%%%%%%%%%%%%%%%%%%
\section{Gravitino production in SUGRA model}
%In this section, we study the consistency between the present new inflation model
%and the SUGRA models.
%As discussed in Ref.~\cite{inflaton}, supergravity effects lead non-trivial mixings 
%between the inflaton and the SUSY breaking fields.
%The mixings between SUSY breaking and inflaton fields result in non-negligible branching ratio
%of inflaton into a pair of gravitinos, which leads non-negligible amount of the gravitino during the %reheating process after inflation.
%Especially, in the case of SUGRA, since we need a SUSY breaking field to be neutral
%under any symmetry, the mixings cannot be suppressed, and the 
%See \cite{Endo:2007sz} as a detailed review of the perturbative inflaton decay into the gravitinos.

The most distinguished  property of the SUGRA models is that they require a SUSY breaking
field which is neutral under any symmetry to obtain gaugino masses of the SSM comparable
to the sfermion masses.
One problem caused by such a singlet SUSY breaking field is so-called Polonyi 
problem~\cite{Coughlan:1983ci,Banks:1993en}
and Polony-induced gravitino problem~\cite{Ibe:2006am}.
In the paper~\cite{Ibe:2006am}, we showed that, thanks to its relatively small Hubble parameter, 
the $R$-invariant New Inflation model is free from the Polonyi problem and the 
Polonyi-induced gravitino 
problem as long as the mass of the SUSY breaking sector field is heavy enough.

The existence of a singlet SUSY breaking field, however, causes another cosmological problem,
that is, the enhancement of the branching ratio of the inflaton into a pair of 
gravitinos~\cite{Kawasaki:2006gs}.
When the SUSY breaking field is a singlet, the mixing between the SUSY breaking field 
and the inflaton after inflation can be enhanced via the supergravity effects.
In our case, the relevant terms which enhance the decay rate of inflaton into a pair of gravitinos
are,
\begin{eqnarray}
\label{eq:Kmix}
 K_{\rm mix} &=& ( C_1^\dagger Z + C_1 Z^\dagger)|\phi|^2 +\cdots, \\
 \label{eq:Wmix}
 W_{\rm mix} &=& C_2 v^2 \phi Z + C_3\frac{g}{5}\phi^5 Z + \cdots,  
\end{eqnarray}
where $Z$ is the SUSY breaking field which has a non-vanihsing $F$-term, $C_{i}$ ($i=1,2,3$) 
constant parameters, and the ellipses the higher dimensional terms.
Since we have no symmetry to suppress the constants $C_{i}$, we naively expect them to be of
the order of one.
Through the supergravity effects, these terms lead to a considerable mixing between
the SUSY breaking field $Z$ and inflaton field $\phi$.

% EFFECTIVE COUPLING
The mixing between the inflaton and the SUSY breaking field leads to an effective coupling of the inflaton 
to gravitinos, $G_{\phi}^{eff}$, with which the decay rate of the inflaton into a pair of the gravitinos
is given by,
\begin{eqnarray}
\label{eq:gamma32}
 \Gamma_{3/2} &=& 
\frac{|G_\phi^{eff}|^2}{288 \pi} \frac{m_\phi^5}{m_{3/2}^2 M_{pl}^2}.
\end{eqnarray}
According to the analysis given in Ref.~\cite{Endo:2006tf}, the effective coupling resulting from 
Eqs.~(\ref{eq:Kmix}) and (\ref{eq:Wmix}) is approximately given by,
\begin{eqnarray}
\label{eq:geff1}
| G_\phi^{eff}|^2 &\simeq& 
3 \vev{\phi}^2 \times
 \left[ C_1 + \frac{1}{16} \left( C_2 + C_3\right) \right]^{2}
\times \left( \frac{m_Z^2}{\mathrm{Max}[m_\phi^2,m_Z^2]}\right)^2.
\end{eqnarray}
Here, $m_{Z}$ denotes the mass of the SUSY breaking field, which is expected to
range from $m_{Z}= O(m_{3/2})$ to $m_{Z} =O( \sqrt{m_{3/2}})$.%
%\footnote{For example, $m_{Z}\simeq m_{3/2}$ is realized in the Polonyi model~\cite{??},
%while it can be enhanced up to $m_{Z}= O(\sqrt{m_{3/2}})$ 
%by a quartic term in the K\"ahler potential in dynamical SUSY breaking models~\cite{??}.
%} 

% REHEATING TEMPERATURE AND YIELD OF THE GRAVITINO
Then, assuming that the inflaton decays mainly into the SSM particles 
with the reheating temperature $T_{R}$, we obtain the gravitino-entropy ratio (yield) 
as,
\begin{eqnarray}
 Y_{3/2}^{\rm inf} &=& 2 \frac{\Gamma_{3/2}}{\Gamma_{R}} \frac{3 T_{R}}{4m_{\phi}},
 \cr
 &\simeq& 4.5 \times |G_\phi^{eff}|^2 
\left(\frac{m_\phi}{10^9\mathrm{GeV}} \right)^4
\left( \frac{1\,{\rm TeV}}{m_{3/2}}\right)^2
\left( \frac{10^7 \mathrm{GeV}}{T_R}\right), \cr
 &\simeq& 2.3 \times 10^{-6} C^2 
 \left(  \frac{\vev\phi}{10^{15}\,{\rm GeV}} \right)^{2}
 \left(\frac{m_\phi}{10^{9}\mathrm{GeV}}\right)^4
 \left( \frac{1\,\rm{TeV}}{m_{3/2}}\right)^2
\left( \frac{10^7 \mathrm{GeV}}{T_R}\right) \cr
& &\times {\rm mim}\left[ m_{Z}^{2}/m_{\phi}^{2},1\right]^{2}, 
\label{eq:y32hiddeninf}
\end{eqnarray}
where $C$ is defined by $C = |C_{1}+(C_{2}+C_{3})/16|$.
In the second equality, we have used Eq.~(\ref{eq:gamma32}) and the relation between
the reheating temperature and the total decay rate of the inflaton $\Gamma_{R}$,
\begin{eqnarray}
\Gamma_{R} =\left(\frac{\pi^{2} g_{*}}{10} \right)^{1/2}T_{R}^{2}.
\end{eqnarray}
Here $g_{*}$ denotes the effective number of massless degrees of freedom during the reheating process, and we use $g_{*}\simeq 230$ which corresponds to the number of the SSM particles.

As we saw in the previous section, the VEV and the mass of the inflaton can be expressed
in terms of the gravitino mass.
Thus, the above yield of the gravitino is determined by three parameters,
the gravitino mass $m_{3/2}$, the mass of the SUSY breaking field $m_{Z}$,
 and the reheating temperature $T_{R}$.
Since the successful BBN requires $T_{R}\lsim 10^{6-7}$\,GeV for $m_{3/2}= O(1)$\,TeV 
to suppress the unstable gravitino abundance produced by the thermal scattering
processes~\cite{Kawasaki:2004qu},
we fix the reheating temperature $T_{R} = 10^{7}$\,GeV in the following of this section.

\begin{figure}[t]
 \begin{minipage}{.50\linewidth}
  \includegraphics[width=.85\linewidth]{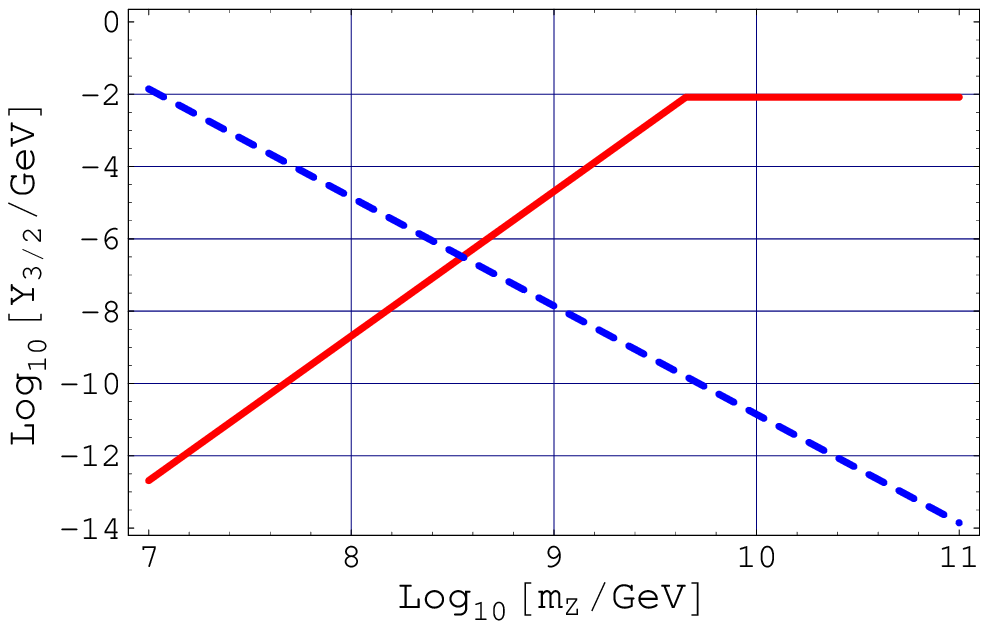}
 \end{minipage}
 \begin{minipage}{.50\linewidth}
  \includegraphics[width=.85\linewidth]{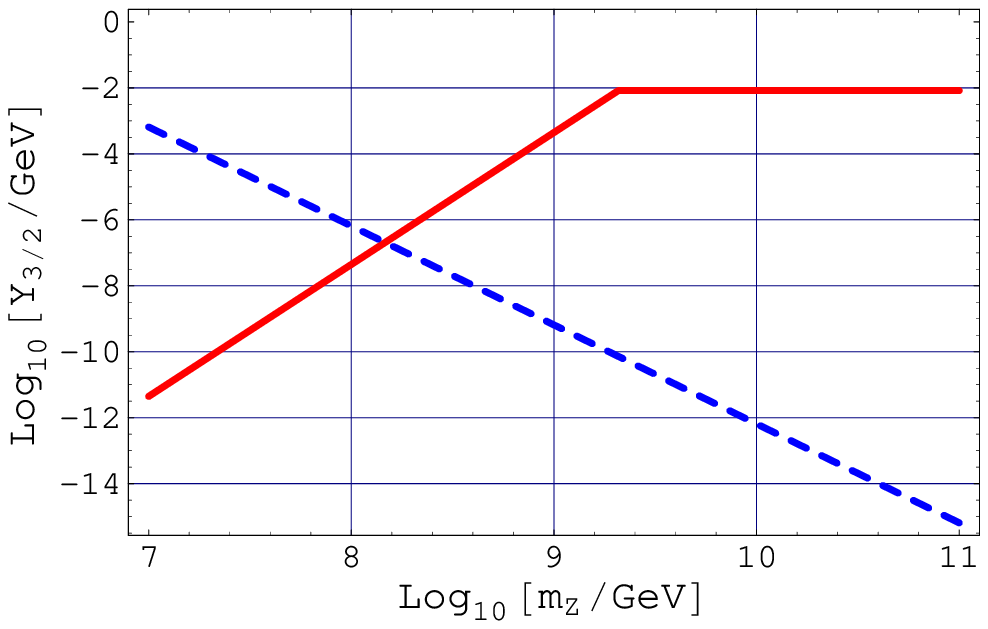}
 \end{minipage}
\caption{The yield of gravitinos for $m_{3/2}=1$\,TeV (left), $100$\,GeV (right).
Solid (red) lines denote  the yields of gravitinos produced by the inflaton decay in 
 Eq.~(\ref{eq:y32hiddeninf}) and dashed (blue)  lines denote the one produced 
by the decay of SUSY breaking field in Eq.~(\ref{eq:y32hiddenpolo}). 
Here, we have taken $C$ and $C_{0}$ to be $1$, and $T_{R}$ to be $10^{7}$\,GeV. }
\label{fig:yieldhidden}
\end{figure}

In Fig.~\ref{fig:yieldhidden}, we show the yield of the gravitino produced by the inflaton decay
as a function of the mass of the SUSY breaking field as solid (red) lines.
As we see from the figure, the yield of the gravitino is suppressed for $m_{Z}<m_{\phi}$,
while $m_{Z}$ dependence disappears for $m_{\phi}<m_{Z}$.

%%%%%%%%%%%%%%%%%%%%%%%%%%%%%%%%%%%%%%%%%%%%%%
%%%%%%%%%%%%%%%%%%%%%%%%%%%%%%%%%%%%%%%%%%%%%%

In order not to spoil the success of the BBN, the gravitino abundance produced by the decay of 
the inflaton must satisfy the constraints in Ref.~\cite{Kawasaki:2004qu,Kohri:2005wn},
\begin{eqnarray}
 Y_{3/2} &\lsim& Y_{3/2}^{\rm upper}, \label{eq:infcon}\\
%Y_{3/2}^{upper} &=&
% \left\{
%  \begin{array}{rcl}
%   1 \times 10^{-16} - 6 \times 10^{-16}& \mathrm{for} & m_{3/2} \simeq 0.1-0.2 \mathrm{TeV}\cr
%   4 \times 10^{-17} - 6 \times 10^{-16}& \mathrm{for} & m_{3/2} \simeq 0.2-2 \mathrm{TeV}\cr
%   7 \times 10^{-17} - 2 \times 10^{-14}& \mathrm{for} & m_{3/2} \simeq 2-10 \mathrm{TeV}\cr
%   6 \times 10^{-13} - 2 \times 10^{-12}& \mathrm{for} & m_{3/2} \simeq 10-30 \mathrm{TeV}\cr
%  \end{array}
%\right. (B_h \simeq 1), \label{eq:hadron1}\\
  Y_{3/2}^{\rm upper} &=&
 \left\{
  \begin{array}{rcl}
   1 \times 10^{-16} - 5 \times 10^{-14}& \mathrm{for} & m_{3/2} \simeq 0.1-1\, \mathrm{TeV}\cr
   2 \times 10^{-14} - 5 \times 10^{-14}& \mathrm{for} & m_{3/2} \simeq 1- 3 \,\mathrm{TeV}\cr
   3 \times 10^{-14} - 2 \times 10^{-13}& \mathrm{for} & m_{3/2} \simeq 3-10\, \mathrm{TeV}\cr
    \end{array}
\right. (B_h \simeq 10^{-3}), \label{eq:hadron2}
\end{eqnarray}
where we have taken the hadronic branching ratio of the gravitino decay to be $B_h\simeq 10^{-3}$ 
for conservative discussion.
The red (solid) lines in Fig.~\ref{fig:yieldhidden} show that we need to require the coefficient $C$
in Eq.~(\ref{eq:y32hiddeninf}) to be very small, i.e. $C\ll 1$ to satisfy the above bound, 
unless the mass of the SUSY breaking field to be much smaller 
than the mass of inflaton, i.e. $m_{Z}\ll m_{\phi}$.

Unfortunately, however, the later option, $m_{Z}\ll m_{\phi}$, brings back the other
cosmological problem, the Polonyi-induced gravitino problem~\cite{Ibe:2006am}. 
Since the SUSY breaking field is neutral under any symmetry, 
there is no reason to forbid  the linear term in the K\"ahler potential,
\begin{equation}
\label{eq:Kshift}
 K_{\rm shift}(Z) = C_0^\dagger Z + C_0 Z^\dagger,
\end{equation}
with the order one coefficient $C_0$.
As discussed in Ref.~\cite{Ibe:2006am},  the above linear term leads to 
a large linear term in the scalar potential of the SUSY breaking field during inflation,
\begin{eqnarray}
\label{eq:hiddenpotential} V(Z) \simeq m_Z^2|Z|^2+3H_{inf}^2(C_0^\dagger Z + C_o Z^\dagger) 
  + \cdots,
\end{eqnarray}
where we have assumed that $H_{\rm inf} \ll m_{Z}$.
By this linear term, the SUSY breaking field is shifted from its VEV by
\begin{equation}
 \vev{Z_{\rm inf}} = \frac{3 H_{inf}^2 C_0}{m_Z^2},
\end{equation}
during inflation,
%
%\footnote{Here, we have defined the origin of $Z$ to be its VEV, although there is no physical meaning
%at $Z =0$. }
 and it begins coherent oscillation after inflation with an amplitude of $O(\vev{Z_{\rm inf}})$.

Once the SUSY breaking field begins its coherent oscillation, it dominantly decays into gravitinos. 
By solving the Boltzmann equation, we obtain the yield of the gravitino from the SUSY breaking field
as,
\begin{eqnarray}
 Y_{3/2}^{\rm hidden} &\simeq& 
 \frac{3}{2}\frac{T_R}{m_{Z}} \frac{m_{Z}^{2}\vev{Z_{\rm inf}}^{2}}{3 H_{\rm inf}^{2}} \cr
   &\simeq& 1.4 \times 10^{-11} {C_0}^{2}
      \left(\frac{m_{3/2}}{1{\rm TeV}}\right)^{4/3}
\left(\frac{10^{10}\mathrm{GeV}}{m_Z}\right)^3
\left( \frac{T_{R}}{10^{7}\,{\rm GeV}} \right),
\label{eq:y32hiddenpolo}
\end{eqnarray}
where we have expressed $H_{\rm inf}$ in terms of $m_{3/2}$ by using the result of the 
previous section.
The dashed (blue) lines in Fig.~\ref{fig:yieldhidden} shows the yield of the gravitino produced by the
decay of the SUSY breaking field for a given gravitino mass.
As we expected, the yield increases when the mass of the SUSY breaking field gets smaller.

As a result, we find that in order not to spoil the success of the BBN the parameters 
$C$ and $C_{0}$ must be suppressed severely, i.e. $C,C_0\lsim 10^{-4}$.
Thus, the $R$-invariant New inflation model with $n=4$ suffers from a fine-tuning problem
to evade the gravitino problem, since we have no reason for such parameters to be suppressed.
Therefore, we find that the $R$-invariant New Inflation model is not successful 
with the SUGRA models with $m_{\rm gaugino} \simeq m_{\rm scalar}\simeq m_{3/2}$.

Before closing this section, we comment on some possible cures of this problem.
It is logically possible to assume that only the gaugino mass terms
break a symmetry under which the SUSY breaking field is charged,
while the terms in Eqs.~(\ref{eq:Kmix}), (\ref{eq:Wmix}) and (\ref{eq:Kshift})
are suppressed at the tree level by the symmetry (see Ref.~\cite{Endo:2007cu}
for related discussion).
In that case, however, the constants $C_0$ and $C_{1}$ are 
generated via at least one- and two-loop diagrams respectively, in which SSM particles circulate.
Thus, the SUGRA models with $m_{\rm gaugino} \simeq m_{\rm scalar}\simeq m_{3/2}$
get $C_{0}=O(0.1)$ and $C_{1}=O(0.01)$ even if they are suppressed at the tree level.
Therefore, if we try to solve the fine-tuning problem by this assumption, we further need to
assume that the gaugino mass is suppressed compared with the scalar masses,
although such hierarchy requires a fine-tunning for the correct electroweak symmetry
breaking.%
\footnote{
The above assumption of  ``soft'' symmetry breaking by only the gaugino mass terms
might work in the large cutoff supergravity proposed in Ref.~\cite{Ibe:2004mp}.
In the large cutoff supergravity, the gaugino masses are suppressed 
compared with the scalar masses while the fine-tuning in the electroweak symmetry breaking 
is not required, thanks to the focus point mechanism~\cite{Barbieri:1987fn,Feng:1999hg,Feng:1999mn}.
}

%%%%%%%%%%%%%%%%%%%%%%%%%%%%%%%%%%%%%%%%%%%%
\section{Gravitino production in Gauge Mediation}
In this section, we consider the GMSB models, where the gravitino 
is the LSP and stable.
As we discussed in the previous section, the gravitino cannot be much lighter 
than $O(10)$\,GeV due to the perturbativity of the inflation model.
Hence, in the following, we concentrate on the case of $m_{3/2}=10$\,GeV as an example.
Besides, we also assume that the SUSY breaking field is charged under some symmetries,
since there is no need to assume it to be neutral in the GMSB models.
(For a neutral SUSY breaking field, 
we have checked that the gravitino abundance produced by the inflaton decay
significantly exceeds 
the observed dark matter abundance for $m_{3/2}=10$\,GeV.)
%As a result, we show that the consistent gravitino dark matter including Baryogenesis scenario 
%can be achieved without significant fine-tuning.

Before going to discuss the gravitino abundance, let us make an assumption about the
reheating process of the new inflation model.
Although there are many possibilities for the reheating mechanism, 
the $R$-invariant New Inflation model has an attractive reheating scenario 
which leads to non-thermal leptogenesis~\cite{Ibe:2006fs}.
By introducing the interaction between the inflaton and the right-handed neutrino $N$'s,
\begin{eqnarray}
\label{eq:neutrino}
W_{\rm neutrino} = \frac{h}{6} \phi^{3}N^{2},
\end{eqnarray}
we can make the inflaton mainly decay into the right-handed neutrino with the reheating
temperature,
\begin{eqnarray}
T_{R} \simeq \left(\frac{10}{g_{*}\pi^{2}} \Gamma_{R}\right)^{1/4}
\simeq 1.5 \times 10^{6}\, h\, g^{-5/4} \,{\rm GeV}. 
\end{eqnarray}
The attractive feature of this reheating process is that the produced right-handed neutrinos
immediately decay into the SSM particles and results in leptogenesis.
In Ref.~\cite{Ibe:2006am,Ibe:2006ck}, we showed that this specific reheating mechanism reproduces 
the observed baryon asymmetry of the universe only for $T_{R}= 10^{6-7}$\,GeV for a
wide range of the parameter $g$.
Thus, for the purpose of finding a cosmologically consistent scenario,
we assume this reheating mechanism with the reheating temperature $T_{R}=10^{6-7}$\,GeV.
We should also mention that this mechanism provides Majorana masses of 
the right-handed neutrinos which is required by the see-saw mechanism~\cite{seesaw}, 
\begin{eqnarray}
m_{N}=\frac{h}{3}\vev{\phi}^{3} \simeq \frac{h}{12 g}m_{\phi}.
\end{eqnarray}

For $m_{3/2}=10$\,GeV and $T_{R}=10^{6-7}$\,GeV, the thermally produced gravitino abundance
is not enough to explain the observed dark matter density as long as 
$m_{\rm gaugino}\leq O(1)$\,TeV~\cite{Moroi:1993mb,Bolz:2000fu,Pradler:2006qh,Rychkov:2007uq}.
Therefore, to explain the observed dark matter density by gravitino, we need to have other sources
of gravitino such as inflaton or the SUSY breaking field as we discussed in the previous section.

First, let us consider the gravitino production from the decay of the SUSY breaking field $Z$.
Notice that there is no linear term in the K\"ahler potential during inflation as in Eq.~(\ref{eq:Kshift}),
since we are assuming that  the SUSY breaking field is charged under some symmetries. 
The dynamics of inflation, however, still shifts the field value of the SUSY breaking field
during inflation via gravitational effect, 
when the SUSY breaking field has a non-vanishing VEV.
That is, during inflation, the SUSY breaking field obtains a so-called Hubble mass term
around its origin,
\begin{eqnarray}
\label{eq:HubbleMass}
 V(Z) \simeq  m_Z^2 |Z- \vev{Z}|^2  +  H_{\rm inf}^2 |Z|^2 + \cdots,
\end{eqnarray}
while it also has a mass term around the VEV $\vev Z$.
Hence, the field value of the SUSY breaking field is shifted from $\vev Z$ by,
\begin{equation}
 \Delta Z = \frac{3 H_{inf}^2}{3H_{inf}^2 + m_Z^2}\vev{Z},
\end{equation}
during inflation.%
\footnote{
A quartic term, $|Z|^{2}|\phi|^{2}$ in the K\"ahler potential changes the coefficient
of the Hubble mass term in Eq.~(\ref{eq:HubbleMass}), although it does not change
our discussion for a wide parameter space. }
Then, as we discussed in the case of SUGRA models, the SUSY breaking field
starts to  oscillate around $\vev Z$ after inflation and produces gravitino when it decays.
By assuming that the SUSY breaking field dominantly decays into gravitinos,%
\footnote{In a class of the GMSB models, the SUSY breaking field can 
dominantly decay into the SSM particles via the interaction for the gauge mediation~\cite{Ibe:2006rc}.
}
we obtain the yield of  gravitinos,
\begin{eqnarray}
 Y_{3/2}^{\rm hidden} 
\simeq 
 \frac{3}{2}\frac{T_R}{m_{Z}} \frac{m_{Z}^{2} \Delta Z^{2}}{3 H_{\rm inf}^{2}}.
\end{eqnarray}
Here, we are assuming that the oscillation of the SUSY breaking field does not dominate the energy 
density of the universe, which is the case for not so large value of $\vev Z$
(see also Ref.~\cite{Ibe:2006rc} for the case where the coherent oscillation dominates the energy density
of the universe).

The above yield of the gravitino is again determined by the gravitino mass, 
the mass of the SUSY breaking field, the reheating temperature and the size of the VEV $\vev{Z}$,
since the Hubble parameter during inflation can be determined for a given gravitino mass.
As discussed above, we take the reheating temperature $T_{R}=10^{6-7}$\,GeV which
is suitable for non-thermal leptogenesis in the $R$-invariant New Inflation model.
As for the VEV of the SUSY breaking field, it is non-trivial to obtain a large VEV while keeping 
the mass of the SUSY breaking field much larger than that of the gravitino.
In our discussion, we take $\vev{Z}=({m_{3/2}^{4} m_{Z}^{-3}})^{1/5}$ as an example by thinking of 
the dynamical SUSY breaking sector discussed in 
Refs.~\cite{Izawa:1995jg,Izawa:1996pk,Intriligator:1996pu} where the SUSY breaking field may have a VEV of the order of the dynamical scale,  $({m_{3/2}^{4} m_{Z}^{-3}})^{1/5}$, 
while the mass of the SUSY breaking field can be high up to $O(\sqrt{m_{3/2}})$.

%%%%%%%%%%%%%%%%%%%%%%%%%%%%%%%%%%%%%%%%%%%%%
\begin{figure}[t]
 \begin{minipage}{.48\linewidth}
  \includegraphics[width=.9\linewidth]{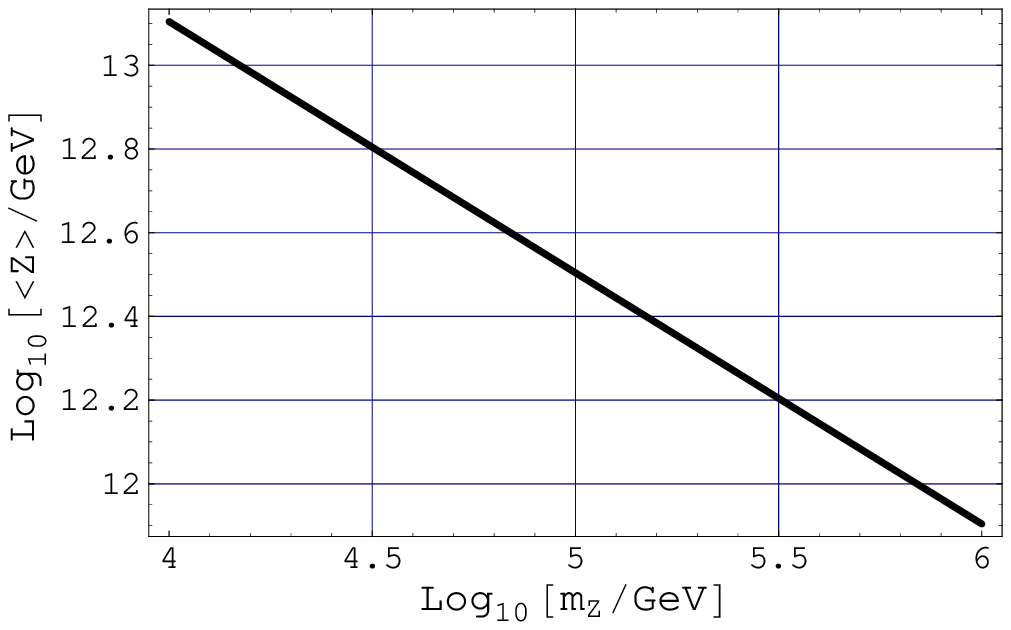}
 \end{minipage}
 \begin{minipage}{.50\linewidth}
  \includegraphics[width=.9\linewidth]{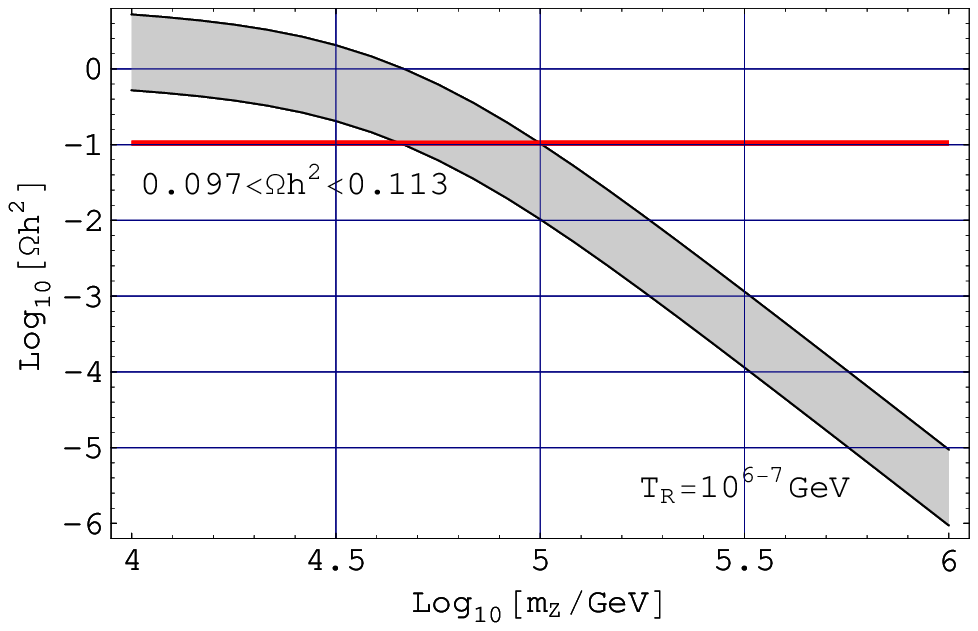}
 \end{minipage}
\caption{Left) The size of the VEV of the SUSY breaking field, 
$\vev{Z} = (m_{3/2}^{4} m_{Z}^{-3})^{1/5}$, which we take as an example.
Right) The mass density parameter of the gravitino dark matter, $\Omega_{3/2}h^2$,  
produced by the decay of SUSY breaking field.
The shaded region corresponds to the density parameter for $T_{R}= 10^{6-7}$\,GeV.
The solid (red) horizontal line shows the observed dark matter density $\Omega h^{2}=0.1050^{+0.0041}_{-0.0040}(1\sigma)$~\cite{Spergel:2006hy}.
In both panels, we have taken $m_{3/2}=10$\,GeV.
}
\label{fig:gmsbpolo}
\end{figure}
%%%%%%%%%%%%%%%%%%%%%%%%%%%%%%%%%%%%%%%%%%%%%

 Altogether, we obtain the mass density paramter of the gravitino as 
\begin{equation}
 \Omega^{\rm hidden}_{3/2}h^2 = 0.1 \times
 \left(\frac{m_{3/2}}{10\,{\rm GeV}} \right)^{7/3}
  \left(\frac{10^{5}\,{\rm GeV}}{m_{Z}}  \right)^{3}
    \left(\frac{T_{R}}{10^{7}\,{\rm GeV}}\right)
  \left(\frac{\vev Z}{10^{12.5}\,{\rm GeV}}\right)^{2},
 \end{equation}
 for $m_{Z}\gg H_{\rm inf}$.
In the right panel of Fig.~\ref{fig:gmsbpolo}, we plot the mass density parameter of the gravitino 
for $\left<Z\right> =({m_{3/2}^{3} m_{Z}^{-1}})^{1/5}$.
We also take $T_{R} = 10^{6-7}$\,GeV for the sake of the baryon asymmetry of the universe.
From the figure, we find that the gravitino produced by the SUSY breaking field can explain 
the observed dark matter density 
$\Omega h^{2}=0.1050^{+0.0041}_{-0.0040}(1\sigma)$~\cite{Spergel:2006hy}. 
Therefore, the $R$-invariant New inflation is not only well consistent with the GMSB models 
 ($m_{3/2}=10$\,GeV), but also naturally provides the dark matter abundance 
and the baryon asymmetry at the same time for a certain parameter range.

Next, let us check that the gravitino abundance produced by the inflaton decay
does not exceed the observed dark matter density.
Since the SUSY breaking field is not neutral, the leading interaction between
the SUSY breaking field and the inflaton in the K\"ahler potential is given by,
\begin{equation}
 K_{int}= b |\phi|^2 |Z|^2.\label{eq:relop}
\end{equation}
In this case, the mixing between SUSY breaking field and the inflaton is much suppressed
compared with the SUGRA models~\cite{Dine:2006ii}.
As a result, the effective coupling of the inflaton to gravitinos is also suppressed.
For $\vev Z = 0$, it is given by,
\begin{eqnarray}
\label{eq:geff2}
 |G_\phi^{eff}|^2 &\simeq& 9 (1-b)^{2}\vev{\phi}^{2}\frac{m_{3/2}^{2}}{m_{\phi}^{2}}
\times\left(\frac{m_{Z}^{2}}{{\rm Max}[m_{Z}^{2},m_{\phi}^{2}]}\right)^{2}.
\end{eqnarray}
Through this effective coupling, the gravitino is produced at the reheating process
of the inflaton.
The resulting mass density parameter of the gravitino is given by,
\begin{eqnarray}
\Omega^{\rm inf}_{3/2} h^2 &=& 2 \times10^{-7}(1-b)^{2}
 \left(  \frac{\vev\phi}{10^{15}\,{\rm GeV}} \right)^{2}
\left(\frac{m_{3/2}}{10\,{\rm GeV}}\right)
\left( \frac{m_{\phi} }{10^9\mathrm{GeV}}\right)^2
\left( \frac{10^6 \mathrm{GeV}}{T_R}\right)\cr
& &\times
{\rm mim}\left[ m_{Z}^{2}/m_{\phi}^{2},1\right]^{2},
\end{eqnarray}
where we have used the yield in the first equality in Eq.~(\ref{eq:y32hiddeninf}).

\begin{figure}[t]
 \begin{center}
  \includegraphics[width=0.45\linewidth ]{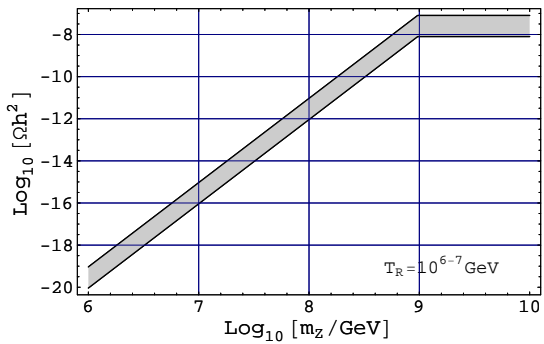}
\caption{The mass density parameter of gravitino  produced by the decay of
  the inflaton for  $m_{3/2}=10{\rm GeV}$, $T_{R} = 10^{6-7}$\,GeV, and $b = 0$.
  For simplicity we have assumed $\vev{Z}= 0$.
   }
\label{fig:gravigmsb} 
 \end{center}
\end{figure}

In Fig.~\ref{fig:gravigmsb}, we plot the gravitino mass density parameter 
from the inflaton decay for $m_{3/2}=10$\,GeV,
$T_{R} = 10^{6-7}$\,GeV, and $b=0$.
From the figure, we see that the gravitino produced by the decay of the inflaton is much smaller
than the observed dark matter density. Thus, the gravitino produced by the inflaton decay with the above specific reheating process 
is subdominant compared with the gravitino produced by the decay of the SUSY breaking field.

For $\vev Z\neq 0$, the above expression of the effective coupling in Eq.~(\ref{eq:geff2})
is changed and becomes complicated.
We have checked, however, that the gravitino dark matter density cannot
be supplied by the decay of the inflaton as long as the VEV of the SUSY breaking 
field is within the order of $({m_{3/2}^{3} m_{Z}^{-1}})^{1/5}$.%
\footnote{
It may be possible to consider a dynamical SUSY breaking model with the VEV much larger than 
we considered here.
In such cases, the gravitino abundance produced by the inflaton decay 
may explain the observed dark matter density.
}

As a result, we find that the $R$-invariant New Inflation model can be consistent with
the GMSB models with $m_{3/2}\simeq 10$\,GeV.
Furthermore, the observed dark matter density can be explained by the gravitino abundance 
produced by the decay of the SUSY breaking field,%
\footnote{
The gravitino abundance produced by the decay of the next to LSP can also contributes 
the dark matter density depending on the details of SSM 
spectrum~\cite{Feng:2003xh,Fujii:2003nr,Ellis:2003dn,Feng:2004zu,Feng:2004mt,Roszkowski:2004jd}, 
which can be a solution to the small scale structure problem of cold dark matter 
cosmology~\cite{Borgani:1996ag,Kaplinghat:2005sy,Cembranos:2005us,Jedamzik:2005sx,Bringmann:2007ft,Feng:2007sx}.
}
while the baryon asymmetry is
provided by the non-termal leptogenesis which is 
naturally embedded into the $R$-invariant New Inflation model.

%%%%%%%%%%%%%%%%%%%%%%%%%%%%%%%%
%%%%%%%%%%%%%%%%%%%%%%%%%%%%%%%%

%%%%%%%%%%%%%%%%%%%%%%%%%%%%%%
\section{Summary}
In this paper, we revisited the $R$-invariant New Inflation model in light of the
recent argument about the gravitino production from the inflaton and the SUSY breaking field. 
As a result, we found that SUGRA models with the $R$-invaraint new inflation model suffer 
from a severe fine-tuning problem where we should have two small parameters of 
$O(10^{-4})$ which are expected to be $O(1)$ without fine-tuning.
On the other hand, we found that the gravitino production from the SUSY breaking field 
is useful in the GMSB models within the reheating temperature which is consistent
with non-thermal leptogenesis.
As we have also shown in Ref.~\cite{Ibe:2006ck}, the gravitino production in the inflaton decay
naturally explains the wino dark matter density in the AMSB models, 
while non-thermal leptogenesis works properly.
Therefore, we conclude that the success of the $R$-invariant New Inflation model 
to predict the spectral index~\cite{Izawa:2003mc,Ibe:2006fs} strongly  suggests 
that the SSM is not realized by the SUGRA where we need a singlet SUSY breaking field,
but by models with mediation mechanisms where we do not require a singlet SUSY breaking
filed, such as  gauge mediation with $m_{3/2}=O(10)$\,GeV 
or anomaly mediation with $m_{3/2}=O(10-100)$\,TeV.
\section*{Acknowledgment}
The authors would like to thank T.~T.~Yanagida for useful discussion at the early stage of the project.
Y. S. thanks the Japan Society for the Promotion of Science for financial support.
This work was supported by the U.S. Department of Energy under contract number 
DE-AC02-76SF00515.


\begin{thebibliography}{99}
%\cite{Chamseddine:1982jx}
\bibitem{Chamseddine:1982jx}
  A.~H.~Chamseddine, R.~Arnowitt and P.~Nath,
%  ``Locally Supersymmetric Grand Unification,''
  Phys.\ Rev.\ Lett.\  {\bf 49}, 970 (1982);
  %%CITATION = PRLTA,49,970;%%
%\cite{Barbieri:1982eh}
%\bibitem{Barbieri:1982eh}
  R.~Barbieri, S.~Ferrara and C.~A.~Savoy,
%  ``Gauge Models with Spontaneously Broken Local Supersymmetry,''
  Phys.\ Lett.\  B {\bf 119}, 343 (1982).
  %%CITATION = PHLTA,B119,343;%%

%\cite{Hall:1983iz}
\bibitem{Hall:1983iz}
  L.~J.~Hall, J.~D.~Lykken and S.~Weinberg,
%  ``Supergravity as the Messenger of Supersymmetry Breaking,''
  Phys.\ Rev.\  D {\bf 27} (1983) 2359.
  %%CITATION = PHRVA,D27,2359;%%

%\cite{Dine:1981za}
\bibitem{Dine:1981za}
  M.~Dine, W.~Fischler and M.~Srednicki,
  %``Supersymmetric Technicolor,''
  Nucl.\ Phys.\ B {\bf 189}, 575 (1981);
  %%CITATION = NUPHA,B189,575;%%
%\cite{Dimopoulos:1981au}
%\bibitem{Dimopoulos:1981au}
  S.~Dimopoulos and S.~Raby,
  ``Supercolor,''
  Nucl.\ Phys.\ B {\bf 192}, 353 (1981);
  %%CITATION = NUPHA,B192,353;%%
%\cite{Dine:1981gu}
%\bibitem{Dine:1981gu}
  M.~Dine and W.~Fischler,
%  ``A Phenomenological Model of Particle Physics Based on Supersymmetry,''
  Phys.\ Lett.\ B {\bf 110}, 227 (1982);
  %%CITATION = PHLTA,B110,227;%%
%\cite{Dine:1982zb}
%\bibitem{Dine:1982zb}
%  M.~Dine and W.~Fischler,
%  ``A Supersymmetric Gut,''
  Nucl.\ Phys.\ B {\bf 204}, 346 (1982);
  %%CITATION = NUPHA,B204,346;%%
%\cite{Nappi:1982hm}
%\bibitem{Nappi:1982hm}
  C.~R.~Nappi and B.~A.~Ovrut,
%  ``Supersymmetric Extension of the SU(3) $\times$ SU(2) $\times$ U(1) Model,''
  Phys.\ Lett.\ B {\bf 113}, 175 (1982);
  %%CITATION = PHLTA,B113,175;%%
%\cite{Alvarez-Gaume:1981wy}
%\bibitem{Alvarez-Gaume:1981wy}
  L.~Alvarez-Gaume, M.~Claudson and M.~B.~Wise,
%  ``Low-Energy Supersymmetry,''
  Nucl.\ Phys.\ B {\bf 207}, 96 (1982);
  %%CITATION = NUPHA,B207,96;%%
%\cite{Dimopoulos:1982gm}
%\bibitem{Dimopoulos:1982gm}
  S.~Dimopoulos and S.~Raby,
%  ``Geometric Hierarchy,''
  Nucl.\ Phys.\ B {\bf 219}, 479 (1983).
  %%CITATION = NUPHA,B219,479;%%


%\cite{Dine:1993yw}
\bibitem{Dine:1993yw}
  M.~Dine and A.~E.~Nelson,
   %``Dynamical supersymmetry breaking at low-energies,''
  %
  Phys.\ Rev.\ D {\bf 48}, 1277 (1993)
  [arXiv:hep-ph/9303230].
  %%CITATION = HEP-PH 9303230;%%

%\cite{Dine:1994vc}
\bibitem{Dine:1994vc}
  M.~Dine, A.~E.~Nelson and Y.~Shirman,
%   ``Low-energy dynamical supersymmetry breaking simplified,''
  %
  Phys.\ Rev.\ D {\bf 51}, 1362 (1995)
  [arXiv:hep-ph/9408384].
  %%CITATION = HEP-PH 9408384;%%

%\cite{Dine:1995ag}
\bibitem{Dine:1995ag}
  M.~Dine, A.~E.~Nelson, Y.~Nir and Y.~Shirman,
   %``New tools for low-energy dynamical supersymmetry breaking,''
  %
  Phys.\ Rev.\ D {\bf 53}, 2658 (1996)
  [arXiv:hep-ph/9507378].
  %%CITATION = HEP-PH 9507378;%%

%\cite{Randall:1998uk}
\bibitem{Randall:1998uk}
  L.~Randall and R.~Sundrum,
%  ``Out of this world supersymmetry breaking,''
  Nucl.\ Phys.\ B {\bf 557}, 79 (1999)
  [arXiv:hep-th/9810155];
  %%CITATION = HEP-TH 9810155;%%
%\cite{Giudice:1998xp}
\bibitem{Giudice:1998xp}
  G.~F.~Giudice, M.~A.~Luty, H.~Murayama and R.~Rattazzi,
%  ``Gaugino mass without singlets,''
  JHEP {\bf 9812}, 027 (1998)
  [arXiv:hep-ph/9810442].
  %%CITATION = HEP-PH 9810442;%%
  
 %%%%%%%%%%% UNSTABLE GRAVITINO %%%%%%%%%%
%\cite{Kawasaki:2004qu}
\bibitem{Kawasaki:2004qu}
  M.~Kawasaki, K.~Kohri and T.~Moroi,
  %``Big-bang nucleosynthesis and hadronic decay of long-lived massive
  %particles,''
 Phys.\ Rev.\  D {\bf 71}, 083502 (2005)
 [arXiv:astro-ph/0408426].
  %%CITATION = PHRVA,D71,083502;%% 
  %\cite{Jedamzik:2006xz}
\bibitem{Jedamzik:2006xz}
  K.~Jedamzik,
  %``Big bang nucleosynthesis constraints on hadronically and
  %electromagnetically decaying relic neutral particles,''
  Phys.\ Rev.\  D {\bf 74}, 103509 (2006)
  [arXiv:hep-ph/0604251].
  %%CITATION = PHRVA,D74,103509;%%
 %%%%%%%%%%% STABLE GRAVITINO %%%%%%%%%%
 %\cite{Moroi:1993mb}
\bibitem{Moroi:1993mb}
  T.~Moroi, H.~Murayama and M.~Yamaguchi,
  %``Cosmological constraints on the light stable gravitino,''
  Phys.\ Lett.\ B {\bf 303}, 289 (1993);
  %%CITATION = PHLTA,B303,289;%%
%\cite{deGouvea:1997tn}
%\bibitem{deGouvea:1997tn}
  A.~de Gouvea, T.~Moroi and H.~Murayama,
  %``Cosmology of supersymmetric models with low-energy gauge mediation,''
  Phys.\ Rev.\ D {\bf 56}, 1281 (1997)
  [arXiv:hep-ph/9701244].
  %%CITATION = HEP-PH 9701244;%%
%\cite{Bolz:2000fu}
\bibitem{Bolz:2000fu}
  M.~Bolz, A.~Brandenburg and W.~Buchmuller,
  %``Thermal production of gravitinos,''
  Nucl.\ Phys.\ B {\bf 606}, 518 (2001)
  [arXiv:hep-ph/0012052].
  %%CITATION = HEP-PH 0012052;%%
  %\cite{Pradler:2006qh}
\bibitem{Pradler:2006qh}
  J.~Pradler and F.~D.~Steffen,
  %``Thermal gravitino production and collider tests of leptogenesis,''
  Phys.\ Rev.\  D {\bf 75}, 023509 (2007)
  [arXiv:hep-ph/0608344];
  %%CITATION = PHRVA,D75,023509;%%
    %\cite{Pradler:2006hh}
%\bibitem{Pradler:2006hh}
%  J.~Pradler and F.~D.~Steffen,
  %``Constraints on the reheating temperature in gravitino dark matter
  %scenarios,''
  Phys.\ Lett.\  B {\bf 648}, 224 (2007)
  [arXiv:hep-ph/0612291].
  %%CITATION = PHLTA,B648,224;%%
  %\cite{Rychkov:2007uq}
\bibitem{Rychkov:2007uq}
  V.~S.~Rychkov and A.~Strumia,
  %``Thermal production of gravitinos,''
  Phys.\ Rev.\  D {\bf 75}, 075011 (2007)
  [arXiv:hep-ph/0701104].
  %%CITATION = PHRVA,D75,075011;%%
%%%%%%%%%%%%%%%%%%%%%%%%%%%%%%%%
%%%%%% MODULI INFLATION %%%%%%%%

%\cite{Endo:2006zj}
\bibitem{Endo:2006zj}
  M.~Endo, K.~Hamaguchi and F.~Takahashi,
%   ``Moduli-induced gravitino problem,''
  %
  Phys.\ Rev.\ Lett.\  {\bf 96}, 211301 (2006)
 [arXiv:hep-ph/0602061].
  %%CITATION = HEP-PH 0602061;
%\cite{Nakamura:2006uc}
\bibitem{Nakamura:2006uc}
  S.~Nakamura and M.~Yamaguchi,
%   ``Gravitino production from heavy moduli decay and cosmological moduli
 %problem revived,''
  Phys.\ Lett.\ B {\bf 638}, 389 (2006)
  [arXiv:hep-ph/0602081].
  %%CITATION = HEP-PH 0602081;%%
%\cite{Dine:2006ii}
\bibitem{Dine:2006ii}
  M.~Dine, R.~Kitano, A.~Morisse and Y.~Shirman,
  %``Moduli decays and gravitinos,''
  Phys.\ Rev.\ D {\bf 73}, 123518 (2006)
 [arXiv:hep-ph/0604140].
  %%CITATION = HEP-PH 0604140;%%
%\cite{Endo:2006tf}
\bibitem{Endo:2006tf}
  M.~Endo, K.~Hamaguchi and F.~Takahashi,
  %``Moduli / inflaton mixing with supersymmetry breaking field,''
  Phys.\ Rev.\ D {\bf 74}, 023531 (2006)
  [arXiv:hep-ph/0605091].
  %%CITATION = HEP-PH 0605091;%%
%\cite{Kawasaki:2006gs}
\bibitem{Kawasaki:2006gs}
  M.~Kawasaki, F.~Takahashi and T.~T.~Yanagida,
  %``Gravitino overproduction in inflaton decay,''
  Phys.\ Lett.\ B {\bf 638}, 8 (2006)
  [arXiv:hep-ph/0603265];
  %%CITATION = HEP-PH 0603265;%%
  %\cite{Kawasaki:2006hm}
%\bibitem{Kawasaki:2006hm}
%  M.~Kawasaki, F.~Takahashi and T.~T.~Yanagida,
  %``The gravitino overproduction problem in inflationary universe,''
  Phys.\ Rev.\ D {\bf 74}, 043519 (2006)
  [arXiv:hep-ph/0605297].
  %%CITATION = HEP-PH 0605297;%%
  %\cite{Asaka:2006bv}
\bibitem{Asaka:2006bv}
  T.~Asaka, S.~Nakamura and M.~Yamaguchi,
  %``Gravitinos from heavy scalar decay,''
  Phys.\ Rev.\ D {\bf 74}, 023520 (2006)
  [arXiv:hep-ph/0604132].
  %%CITATION = HEP-PH 0604132;%%
    %\cite{Endo:2006qk}
\bibitem{Endo:2006qk}
  M.~Endo, M.~Kawasaki, F.~Takahashi and T.~T.~Yanagida,
  %``Inflaton decay through supergravity effects,''
  Phys.\ Lett.\ B {\bf 642}, 518 (2006)
  [arXiv:hep-ph/0607170].
  %%CITATION = HEP-PH 0607170;%%
 %\cite{Endo:2006xg}
\bibitem{Endo:2006xg}
  M.~Endo, K.~Kadota, K.~A.~Olive, F.~Takahashi and T.~T.~Yanagida,
  %``The decay of the inflaton in no-scale supergravity,''
  JCAP {\bf 0702}, 018 (2007)
  [arXiv:hep-ph/0612263].
  %%CITATION = JCAPA,0702,018;%%  
 %\cite{Endo:2007ih}
\bibitem{Endo:2007ih}
  M.~Endo, F.~Takahashi and T.~T.~Yanagida,
  %``Anomaly-induced inflaton decay and gravitino-overproduction problem,''
  arXiv:hep-ph/0701042;
  %%CITATION = HEP-PH/0701042;%%
%\cite{Endo:2007sz}
%\bibitem{Endo:2007sz}
  M.~Endo, F.~Takahashi and T.~T.~Yanagida,
  %``Inflaton Decay in Supergravity,''
  arXiv:0706.0986 [hep-ph].
  %%CITATION = ARXIV:0706.0986;%%
   %%%%%%%%%%%%%%%%%%%%%%%%%%%%%
   %% R-invariant Inflation %%%%
   %\cite{Kumekawa:1994gx}
\bibitem{Kumekawa:1994gx}
  K.~Kumekawa, T.~Moroi and T.~Yanagida,
  %``Flat potential for inflaton with a discrete R invariance in supergravity,''
  Prog.\ Theor.\ Phys.\  {\bf 92}, 437 (1994)
  [arXiv:hep-ph/9405337].
  %%CITATION = PTPKA,92,437;%%
%\cite{Izawa:1996dv}
\bibitem{Izawa:1996dv}
  K.~I.~Izawa and T.~Yanagida,
  %``Natural new inflation in broken supergravity,''
  Phys.\ Lett.\  B {\bf 393}, 331 (1997)
  [arXiv:hep-ph/9608359].
  %%CITATION = PHLTA,B393,331;%%
  %\cite{Izawa:2003mc}
\bibitem{Izawa:2003mc}
  K.~I.~Izawa,
  %``Supergravity minimal inflation and its spectral index revisited,''
  Phys.\ Lett.\  B {\bf 576}, 1 (2003)
  [arXiv:hep-ph/0305286].
  %%CITATION = PHLTA,B576,1;%%
  %\cite{Ibe:2006fs}
\bibitem{Ibe:2006fs}
  M.~Ibe, K.~I.~Izawa, Y.~Shinbara and T.~T.~Yanagida,
  %``Minimal supergravity, inflation, and all that,''
  Phys.\ Lett.\  B {\bf 637}, 21 (2006)
  [arXiv:hep-ph/0602192].
  %%CITATION = PHLTA,B637,21;%%
 
 
 %%%%%%%%%%%%%%%%%%%%%%%%%%%%%%%%%%%%%
%\cite{Spergel:2006hy}
\bibitem{Spergel:2006hy}
  D.~N.~Spergel {\it et al.}  [WMAP Collaboration],
  %``Wilkinson Microwave Anisotropy Probe (WMAP) three year results:
  %Implications for cosmology,''
  arXiv:astro-ph/0603449.
  %%CITATION = ASTRO-PH/0603449;%%
%\cite{Fukugita:1986hr}

%%%%%%%%%%%%%%%%%%%%%%%%%%%%%%%%%%%%%
\bibitem{Fukugita:1986hr}
  M.~Fukugita and T.~Yanagida,
  %``Baryogenesis Without Grand Unification,''
  Phys.\ Lett.\  B {\bf 174}, 45 (1986); \\
 For a review,
  %%CITATION = PHLTA,B174,45;%%
  W.~Buchmuller, R.~D.~Peccei and T.~Yanagida,
  %``Leptogenesis as the origin of matter,''
  Ann.\ Rev.\ Nucl.\ Part.\ Sci.\  {\bf 55}, 311 (2005)
  [arXiv:hep-ph/0502169].
  %%CITATION = ARNUA,55,311;%%



 %%%%%%%%%%%%%%%%%%%%%%%%%%%%%%%%%%
%\cite{Campbell:1992hd}
\bibitem{Campbell:1992hd}
  B.~A.~Campbell, S.~Davidson and K.~A.~Olive,
  %``Inflation, neutrino baryogenesis, and (S)neutrino induced baryogenesis,''
  Nucl.\ Phys.\  B {\bf 399}, 111 (1993)
  [arXiv:hep-ph/9302223].
  %%CITATION = NUPHA,B399,111;%%
  %\cite{Asaka:1999yd}
\bibitem{Asaka:1999yd}
  T.~Asaka, K.~Hamaguchi, M.~Kawasaki and T.~Yanagida,
  %``Leptogenesis in inflaton decay,''
  Phys.\ Lett.\  B {\bf 464}, 12 (1999)
  [arXiv:hep-ph/9906366];
  %%CITATION = PHLTA,B464,12;%%
  %\cite{Asaka:1999jb}
%\bibitem{Asaka:1999jb}
%  T.~Asaka, K.~Hamaguchi, M.~Kawasaki and T.~Yanagida,
  %``Leptogenesis in inflationary universe,''
  Phys.\ Rev.\  D {\bf 61}, 083512 (2000)
  [arXiv:hep-ph/9907559].
  %%CITATION = PHRVA,D61,083512;%%
   %\cite{Ibe:2005jf}
  \bibitem{Ibe:2005jf}
  M.~Ibe, T.~Moroi and T.~Yanagida,
  %``Dark matter and baryon asymmetry of the universe in large-cutoff
  %supergravity,''
  Phys.\ Lett.\  B {\bf 620}, 9 (2005)
  [arXiv:hep-ph/0502074].
  %%CITATION = PHLTA,B620,9;%%

%%%%%%%%%%%%%%%%%%%%%%%%%%%%%%%%%
%\cite{Ibe:2006am}
\bibitem{Ibe:2006am}
  M.~Ibe, Y.~Shinbara and T.~T.~Yanagida,
  %``The Polonyi problem and upper bound on inflation scale in supergravity,''
  Phys.\ Lett.\  B {\bf 639}, 534 (2006)
  [arXiv:hep-ph/0605252].
  %%CITATION = PHLTA,B639,534;%%
  

%\cite{Coughlan:1983ci}
\bibitem{Coughlan:1983ci}
  G.~D.~Coughlan, W.~Fischler, E.~W.~Kolb, S.~Raby and G.~G.~Ross,
  %``Cosmological Problems For The Polonyi Potential,''
  Phys.\ Lett.\ B {\bf 131}, 59 (1983).
  %%CITATION = PHLTA,B131,59;%%

%\cite{Banks:1993en}
\bibitem{Banks:1993en}
  T.~Banks, D.~B.~Kaplan and A.~E.~Nelson,
  %``Cosmological implications of dynamical supersymmetry breaking,''
  Phys.\ Rev.\ D {\bf 49}, 779 (1994)
  [arXiv:hep-ph/9308292].
  %%CITATION = HEP-PH 9308292;%%
  
\bibitem{Kohri:2005wn}
  K.~Kohri, T.~Moroi and A.~Yotsuyanagi,
  %``Big-bang nucleosynthesis with unstable gravitino and upper bound on the
  %reheating temperature,''
  Phys.\ Rev.\  D {\bf 73} (2006) 123511
  [arXiv:hep-ph/0507245].
  %%CITATION = PHRVA,D73,123511;%%

  %\cite{Endo:2007cu}
\bibitem{Endo:2007cu}
  M.~Endo, F.~Takahashi and T.~T.~Yanagida,
  %``Retrofitted gravity mediation without the gravitino overproduction
  %problem,''
  arXiv:hep-ph/0702247.
  %%CITATION = HEP-PH/0702247;%%
  
  %\cite{Ibe:2004mp}
\bibitem{Ibe:2004mp}
  M.~Ibe, K.~I.~Izawa and T.~Yanagida,
  %``Realization of minimal supergravity,''
  Phys.\ Rev.\  D {\bf 71}, 035005 (2005)
  [arXiv:hep-ph/0409203].
  %%CITATION = PHRVA,D71,035005;%%
%\cite{Feng:1999mn}
%\cite{Barbieri:1987fn}
\bibitem{Barbieri:1987fn}
  R.~Barbieri and G.~F.~Giudice,
  %``Upper Bounds On Supersymmetric Particle Masses,''
  Nucl.\ Phys.\  B {\bf 306}, 63 (1988).
  %%CITATION = NUPHA,B306,63;%%
%\cite{Feng:1999hg}
\bibitem{Feng:1999hg}
  J.~L.~Feng and T.~Moroi,
  %``Supernatural supersymmetry: Phenomenological implications of
  %anomaly-mediated supersymmetry breaking,''
  Phys.\ Rev.\  D {\bf 61}, 095004 (2000)
  [arXiv:hep-ph/9907319].
  %%CITATION = PHRVA,D61,095004;%%
\bibitem{Feng:1999mn}
  J.~L.~Feng, K.~T.~Matchev and T.~Moroi,
  %``Multi-TeV scalars are natural in minimal supergravity,''
  Phys.\ Rev.\ Lett.\  {\bf 84}, 2322 (2000)
  [arXiv:hep-ph/9908309];
  %%CITATION = PRLTA,84,2322;%%
  %\cite{Feng:1999zg}
%\bibitem{Feng:1999zg}
%  J.~L.~Feng, K.~T.~Matchev and T.~Moroi,
  %``Focus points and naturalness in supersymmetry,''
  Phys.\ Rev.\  D {\bf 61}, 075005 (2000)
  [arXiv:hep-ph/9909334].
  %%CITATION = PHRVA,D61,075005;%%

\bibitem{seesaw}
  T. Yanagida,
  in \textit{Proc. Workshop on the Unified Theory and Baryon Number in the Universe},
  ed. by O. Sawada, A. Sugamoto
  (KEK report 79-18, 1979), p. 95;
  M. Gell-Mann, P. Ramond, R. Slansky,
  in \textit{Supergravity},
  ed. by P. van Nieuwenhuizen, D.Z. Freedman
  (North Holland, Amsterdam 1979), p. 315.




  
%\cite{Ibe:2006rc}
\bibitem{Ibe:2006rc}
  M.~Ibe and R.~Kitano,
  %``Gauge mediation in supergravity and gravitino dark matter,''
  Phys.\ Rev.\  D {\bf 75}, 055003 (2007)
  [arXiv:hep-ph/0611111];
  %%CITATION = PHRVA,D75,055003;%%
  %\cite{Ibe:2007km}
%\bibitem{Ibe:2007km}
  M.~Ibe and R.~Kitano,
  %``Sweet Spot Supersymmetry,''
  arXiv:0705.3686 [hep-ph].
  %%CITATION = ARXIV:0705.3686;%%


%%%%%%%%%%%%%%%%%%%%%%%%%%%%%%%%%%%
%\cite{Izawa:1995jg}
\bibitem{Izawa:1995jg}
  K.~I.~Izawa and T.~Yanagida,
  %``QCD - Like Hidden Sector Models Without The Polonyi Problem,''
  Prog.\ Theor.\ Phys.\  {\bf 94}, 1105 (1995)
  [arXiv:hep-ph/9507441].
  %%CITATION = PTPKA,94,1105;%%
%\cite{Izawa:1996pk}
\bibitem{Izawa:1996pk}
  K.~I.~Izawa and T.~Yanagida,
  %``Dynamical Supersymmetry Breaking in Vector-like Gauge Theories,''
  Prog.\ Theor.\ Phys.\  {\bf 95}, 829 (1996)
  [arXiv:hep-th/9602180].
  %%CITATION = PTPKA,95,829;%%
  %\cite{Intriligator:1996pu}
\bibitem{Intriligator:1996pu}
  K.~A.~Intriligator and S.~D.~Thomas,
  %``Dynamical Supersymmetry Breaking on Quantum Moduli Spaces,''
  Nucl.\ Phys.\  B {\bf 473}, 121 (1996)
  [arXiv:hep-th/9603158].
  %%CITATION = NUPHA,B473,121;%%
  %%%%%%%%%%%%%%%%%%%%%%%%%%%%%%%%%
    
%\cite{Ibe:2006ck}
\bibitem{Ibe:2006ck}
  M.~Ibe, Y.~Shinbara and T.~T.~Yanagida,
  %``A new inflation model with anomaly-mediated supersymmetry breaking,''
  Phys.\ Lett.\  B {\bf 642}, 165 (2006)
  [arXiv:hep-ph/0608127].
  %%CITATION = PHLTA,B642,165;%%

%%%%%%%%%%%%%%%%%%%%%%%%%%%%%  
  %\cite{Feng:2003xh}
\bibitem{Feng:2003xh}
  J.~L.~Feng, A.~Rajaraman and F.~Takayama,
  %``Superweakly-interacting massive particles,''
  Phys.\ Rev.\ Lett.\  {\bf 91}, 011302 (2003)
  [arXiv:hep-ph/0302215];
  %%CITATION = PRLTA,91,011302;%%
  %\cite{Feng:2003uy}
%\bibitem{Feng:2003uy}
%  J.~L.~Feng, A.~Rajaraman and F.~Takayama,
  %``SuperWIMP dark matter signals from the early universe,''
  Phys.\ Rev.\  D {\bf 68}, 063504 (2003)
  [arXiv:hep-ph/0306024].
  %%CITATION = PHRVA,D68,063504;%%
  %\cite{Fujii:2003nr}
\bibitem{Fujii:2003nr}
  M.~Fujii, M.~Ibe and T.~Yanagida,
  %``Upper bound on gluino mass from thermal leptogenesis,''
  Phys.\ Lett.\  B {\bf 579}, 6 (2004)
  [arXiv:hep-ph/0310142].
  %%CITATION = PHLTA,B579,6;%%
%\cite{Ellis:2003dn}
\bibitem{Ellis:2003dn}
  J.~R.~Ellis, K.~A.~Olive, Y.~Santoso and V.~C.~Spanos,
  %``Gravitino dark matter in the CMSSM,''
  Phys.\ Lett.\  B {\bf 588}, 7 (2004)
  [arXiv:hep-ph/0312262].
  %%CITATION = PHLTA,B588,7;%%
  %\cite{Feng:2004zu}
\bibitem{Feng:2004zu}
  J.~L.~Feng, S.~f.~Su and F.~Takayama,
  %``SuperWIMP gravitino dark matter from slepton and sneutrino decays,''
  Phys.\ Rev.\  D {\bf 70}, 063514 (2004)
  [arXiv:hep-ph/0404198].
  %%CITATION = PHRVA,D70,063514;%%
  %\cite{Feng:2004mt}
\bibitem{Feng:2004mt}
  J.~L.~Feng, S.~Su and F.~Takayama,
  %``Supergravity with a gravitino LSP,''
  Phys.\ Rev.\  D {\bf 70}, 075019 (2004)
  [arXiv:hep-ph/0404231].
  %%CITATION = PHRVA,D70,075019;%%
%\cite{Roszkowski:2004jd}
\bibitem{Roszkowski:2004jd}
  L.~Roszkowski, R.~Ruiz de Austri and K.~Y.~Choi,
  %``Gravitino dark matter in the CMSSM and implications for leptogenesis  and
  %the LHC,''
  JHEP {\bf 0508}, 080 (2005)
  [arXiv:hep-ph/0408227].
  %%CITATION = JHEPA,0508,080;%%
%%%%%%%%%%%%%%%%%%%%%%%%%%%%%%  
  %\cite{Borgani:1996ag}
\bibitem{Borgani:1996ag}
  S.~Borgani, A.~Masiero and M.~Yamaguchi,
  %``Light gravitinos as mixed dark matter,''
  Phys.\ Lett.\  B {\bf 386}, 189 (1996)
  [arXiv:hep-ph/9605222].
  %%CITATION = PHLTA,B386,189;%%
  %\cite{Kaplinghat:2005sy}
\bibitem{Kaplinghat:2005sy}
  M.~Kaplinghat,
  %``Dark matter from early decays,''
  Phys.\ Rev.\  D {\bf 72}, 063510 (2005)
  [arXiv:astro-ph/0507300].
  %%CITATION = PHRVA,D72,063510;%%
  %\cite{Cembranos:2005us}
\bibitem{Cembranos:2005us}
  J.~A.~R.~Cembranos, J.~L.~Feng, A.~Rajaraman and F.~Takayama,
  %``SuperWIMP solutions to small scale structure problems,''
  Phys.\ Rev.\ Lett.\  {\bf 95}, 181301 (2005)
  [arXiv:hep-ph/0507150].
  %%CITATION = PRLTA,95,181301;%%
  %\cite{Jedamzik:2005sx}
\bibitem{Jedamzik:2005sx}
  K.~Jedamzik, M.~Lemoine and G.~Moultaka,
  %``Gravitino, axino, Kaluza-Klein graviton warm and mixed dark matter and
  %reionisation,''
  JCAP {\bf 0607}, 010 (2006)
  [arXiv:astro-ph/0508141].
  %%CITATION = JCAPA,0607,010;%%
  %\cite{Bringmann:2007ft}
\bibitem{Bringmann:2007ft}
  T.~Bringmann, F.~Borzumati and P.~Ullio,
  %``Dark matter from late decays and the small-scale structure problems,''
  arXiv:hep-ph/0701007.
  %%CITATION = HEP-PH/0701007;%%
  %\cite{Feng:2007sx}
\bibitem{Feng:2007sx}
  J.~L.~Feng, B.~T.~Smith and F.~Takayama,
  %``Goldilocks Supersymmetry,''
  arXiv:0709.0297 [hep-ph].
  %%CITATION = ARXIV:0709.0297;%%


  
\end{thebibliography}
\end{document}